# Arbitrary non-paraxial accelerating periodic beams and spherical shaping of light

A. Mathis, F. Courvoisier*, R. Giust, L. Furfaro, M. Jacquot, L. Froehly, J. M. Dudley

*Département d'Optique P.M. Duffieux, Institut FEMTO-ST,*
*UMR 6174 CNRS Université de Franche-Comté, 25030 Besançon Cedex, France*
*\*Corresponding author: francois.courvoisier@femto-st.fr*

We report the observation of arbitrary accelerating beams designed using a non-paraxial description of optical caustics. We use a spatial light modulator-based setup and techniques of Fourier optics to generate circular and Weber beams subtending over 95 degrees of arc. Applying a complementary binary mask also allows the generation of periodic accelerating beams taking the forms of snake-like trajectories, and the application of a rotation to the caustic allows the first experimental synthesis of optical accelerating beams upon the surface of a sphere in three dimensions.

Accelerating optical beams are a novel class of electromagnetic wave associated with a localized intensity maximum that propagates along a curved trajectory. Since initial work studying Airy beam solutions of the paraxial wave equation propagating along parabolic trajectories [1], more general classes of solutions have been obtained including Mathieu beams along elliptical trajectories and Weber beams along parabolic trajectories [2-4].

Approaches used to find accelerating beam solutions have varied. Deriving exact solutions from Maxwell's equations or from the derived Helmholtz wave equation is the most general approach, and yields significant theoretical insight [5]. However, other studies have investigated approaches to engineer an incident beam such that its subsequent propagation evolves on a pre-defined and *arbitrary* acceleration trajectory [6, 7].

A key concept in the design of arbitrary accelerating beams is the association of the target trajectory with an optical caustic [6, 8]. Previous results using caustic design for non-paraxial accelerating beam generation, however, have been associated with non uniform illumination and small arc segments because phase modification to generate the caustic has been applied directly to the incident field [9].

In this paper we describe a non-paraxial technique for accelerating beam design in the Fourier domain that overcomes existing limitations and which can generate arbitrary accelerating beams in two and three dimensions with uniform illumination over more than a quadrant of a circle. We report experimental synthesis of beams following parabolic (Weber), circular and quartic acceleration trajectories. We also show that our method of non-paraxial beam design is compatible with binary modulation and the generation of periodic (snake-like) propagation. Moreover, phase mask rotation extends the generation of non-paraxial circular beams into three dimensions, allowing us to also report the first observation of an accelerating wave generated to propagate along the surface of a sphere [10].

We first note that although it is often believed that caustic descriptions are limited to geometrical optics, provided one considers scalar unidirectional propagation they can be straightforwardly included in wave optics through diffraction integral theory, without any paraxial approximation [9]. Considering firstly a two dimensional fold caustic propagating along z and accelerating along y we wish to determine the generating (Fourier) phase profile $\Phi_c(y_F)$ associated with a target trajectory $c(z)$ in the focus of a large numerical aperture microscope objective (MO) acting as a Fourier transform lens. The setup and coordinate systems are shown in Fig. 1. Using a Debye-Wolf diffraction description of focusing by the MO [11], we obtain the parametric description of the generating phase as follows:

$$\frac{\partial \Phi_c(y_F)}{\partial y_F} = k \frac{c(z) - z \partial_z c(z)}{f} \quad ; y_F = f \frac{\partial_z c(z)}{\sqrt{1 + \left[\partial_z c(z)\right]^2}} \quad (1)$$

with $k$ the wave number and $f$ the MO focal length. Although Eq. (1) does not yield the intensity distribution of the accelerating beam, it nonetheless describes the generating phase profile $\Phi_c(y_F)$ for completely arbitrary trajectories. In fact, we have verified that it yields phase profiles identical to the generating profiles found from Maxwell's equations for circular [12] and Weber beams [3] where analytical integration is possible, and in the more general case for arbitrary $c(z)$, numerical integration can be used.

Once the phase profile in the plane before the MO is known, it is straightforward to numerically determine the phase mask to apply to a spatial light modulator (SLM) to effect the necessary transformation to an incident beam, taking into account any subsequent imaging and/or beam reduction optics.

Fig. 1 shows the experimental setup. A 632.8 nm laser is phase-modulated by a LCOS-Hamamatsu SLM where the laser beam fills the SLM aperture. The SLM is imaged by a $4f$ telescope (L1, L2) onto the back focal plane of a × 50 ($f$ = 3.6 mm) 0.8 numerical aperture MO, filling the entrance pupil. Before the MO we use an initial stage of pre-filtering using Fourier optics to provide additional flexibility, with spatial filtering of the zeroth order from the SLM. The phase function is encoded along $y_F$ direction and the input beam is polarized along $x_F$. The beam is imaged by translating a × 50 microscope objective before fixed lens L3 and a CCD camera.

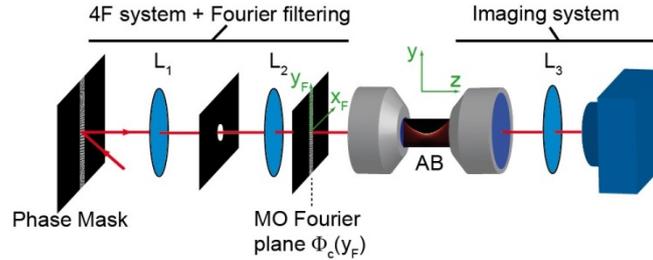

Fig. 1. Experimental setup. Target accelerating beam (AB) is generated in the Fourier plane of a microscope objective by applying a phase-mask applied to a spatial light modulator (in first order of diffraction). L1, L2, and L3 are lenses of focal lengths 500, 250 and 200 mm respectively.

Fig. 2 shows experimental results for three different non-paraxial trajectories: parabolic $c(z) = az^2$ with $a = 0.006 \mu m^{-1}$; circular with radius R = 75 µm; and quartic $c(z) = az^4$ with $a = 6 \times 10^{-5} \mu m^{-3}$. The left column shows the measured intensity distribution compared to the defined target trajectories (white dashed lines). All intensity distributions are shown with a 1:1 aspect ratio.

Before performing experiments, we numerically determined the expected intensity profiles by propagating a phase-modulated incident gaussian beam through our system (including all effects such as SLM sampling, finite beam size, etc) using the angular spectrum of plane waves method and the integration of the Debye integral of each field component [11]. These numerical results are essentially indistinguishable from experiment and are not shown. To highlight the excellent agreement between the target and experimental trajectories, the adjacent subfigures show expanded details of the intensity distributions at the points A and B as shown. The experimentally realized beams are nonparaxial, subtending >95° of arc at intensities >0.5% of the maximal intensity.

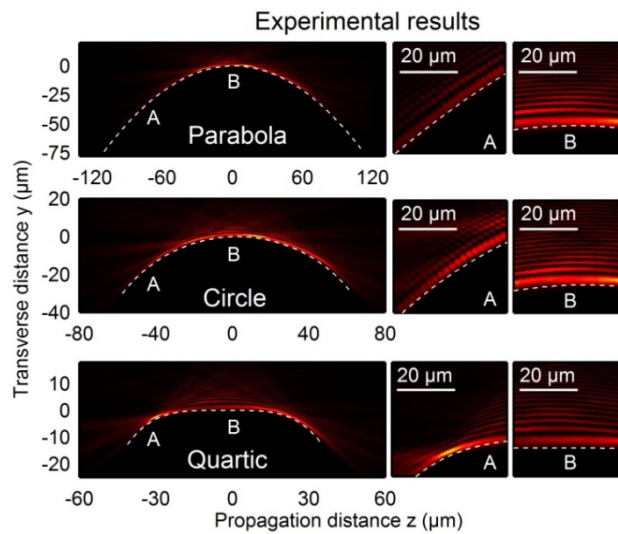

Fig. 2. Experimental intensity distributions: (a) parabolic Weber beam, (b) circular Mathieu beam (c) quartic beam. Adjacent subfigures in each case show expanded views at points A and B. The white dashed lines show the target trajectories. Vertical and horizontal scales are identical.

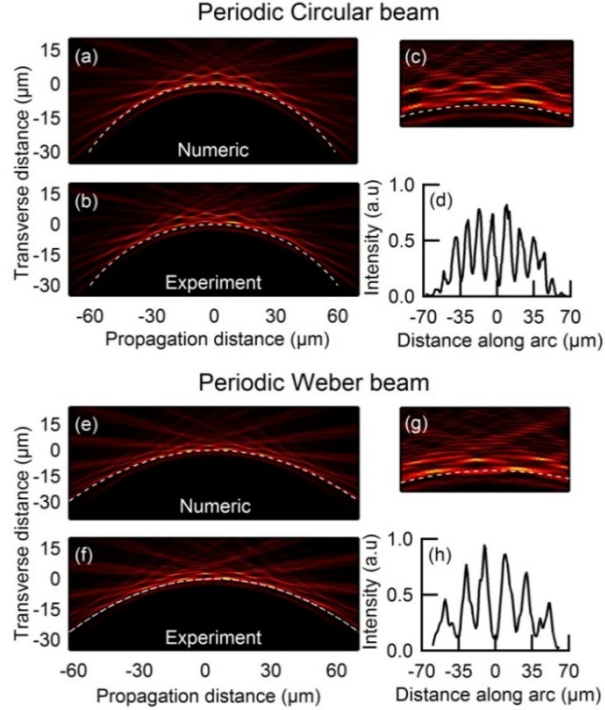

**Fig. 3.** Periodic circular and parabolic beams. Dashed lines show target trajectories. Vertical and horizontal scales are identical. (a,e) numerical results. (b,f) Experimental results. (c,g) Detailed view near trajectory extremum (d,h) Experimental line profiles along the caustics.

An important feature of our experimental setup is that the Fourier filtering after lens L1 allows phase modulation with the SLM to also encode amplitude modulation. This is because the combination of phase modulation and Fourier filtering can spatially modify the diffraction efficiency such that light not diffracted into the first order is sent into the zero order, effectively leading to amplitude modulation in the MO Fourier plane [13]. This allows flexible amplitude control along the length of the accelerating trajectory, by combining the phase profile on the SLM with an additional binary phase mask $M(y_F)$ such that the intensity profile at the focal region of the MO is $I(y,z) \propto |A_c(y,z) * m(y)|^2$ where $A_c(y,z)$ is the field amplitude, $m(y)$ and $M(y_F)$ are Fourier transform pairs, and $*$ denotes convolution product.

This approach provides a supplementary degree of freedom to modulate the field amplitude of the accelerating beam. Periodically modulated accelerating beams can equivalently be interpreted physically in terms of the beating between the superposition of accelerating beams with different propagation constants. Indeed it was this concept that motivated their first experimental realization using two superposed circular beams [12]. The approach here is effectively associated with the superposition of a larger number of accelerating beam components generated from the binary modulation function, providing a general framework for their experimental implementation at high contrast and with any accelerating trajectory.

Experimental results for both circular and parabolic beams are shown in Fig. 3 (a-d) for circular and Fig. 3 (e-h) for parabolic periodic trajectories. In this case we also show numerical results for the expected trajectories which were used to confirm the phase mask design, and we see very good agreement. The detailed views in (c) and (g) highlight clearly the snake-like beam structure [12] and the line profiles in (d) and (h) show the high contrast along the caustic.

The ability to generate non-paraxial accelerating beams over more than a full quadrant of a circle with our setup also opens up the possibility to generate spherical optical fields, a recently-introduced class of shape-preserving solution to Maxwell's equations [10]. This is implemented by combining the phase mask of a circular caustic as described above with an additional rotational phase over 180° corresponding to a transverse radius identical to that of the caustic propagating along z as shown in the inset in Fig. 4 (a). This then yields the generation of a 3-dimensional caustic sheet upon the surface of a sphere.

Experimental results are shown in Fig. 4. Although the rotational phase can yield a full half-circle in the transverse xy plane, the longitudinal extent of the caustic is limited by the MO numerical aperture to 104° of arc. The generated field in this case takes the form of a spherical cap extended down to the equator. Figs. 4(a) and (b) show views of this iso-intensity surface at 30% of maximal intensity, clearly showing the remarkable structure of this field. Figs 4(c) and (d) show xz and xy cross sections respectively.

There are several important conclusions to be drawn from this work. The use of non-paraxial Debye Wolf wave diffraction theory allows the design and experimental generation of arbitrary non-paraxial beams over arc angles exceeding 90° and where the experimental trajectory is in excellent agreement with the calculated target caustic. The additional amplitude modulation of the generating phase profile has been shown to yield high contrast periodic accelerating beams, again over non-paraxial angles of more than 90°. Finally, rotation of an accelerating beam over 180° in the transverse plane has demonstrated the first experimental generation of caustic propagation upon a spherical surface. We anticipate that our results will motivate significant further studies of accelerating beams in 2 and 3 dimensions in the non-paraxial regime.

We thank the Région of Franche-Comté and the French ANR, contract 2011-BS04-010-01 NANOFLAM. This work has been performed in cooperation with the Labex ACTION program (contract ANR-11-LABX-01-01).

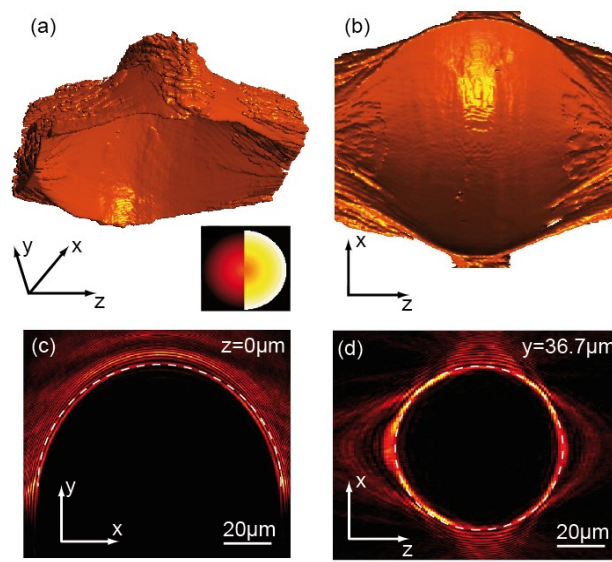

**Fig. 4.** Experimental spherically shaped field. (a) and (b) 3D isointensity surfaces at 30% of maximal intensity. Inset: rotation phase mask (unwrapped). (c) and (d) Cross-sections in xy and xz planes. White dashed lines show the cross sections of a 50 μm radius half-sphere. Scaling is identical in all directions.

1. G. A. Siviloglou, J. Broky, A. Dogariu and D. N. Christodoulides, Phys. Rev. Lett. **99**, 213901 (2007).
2. I. Kaminer, R. Bekenstein, J. Nemirovsky and M. Segev, Phys. Rev. Lett. **108**, 163901 (2012).
3. P. Zhang, Y. Hu, T. Li, D. Cannan, X. Yin, R. Morandotti, Z. Chen and X. Zhang, Phys. Rev. Lett. **109**, 193901 (2012).
4. M. A. Bandres and B. M. Rodriguez-Lara, New J. Phys. **15**, 013054 (2013).
5. P. Aleahmad, M. A. Miri, M. S. Mills, I. Kaminer, M. Segev and D. N. Christodoulides, Phys. Rev. Lett. **109**, 203902 (2012).
6. L. Froehly, F. Courvoisier, A. Mathis, M. Jacquot, L. Furfaro, R. Giust, P. A. Lacourt and J. M. Dudley, Opt. Express **19**, 16455 (2011).
7. E. Greenfield, M. Segev, W. Walasik, and O. Raz, Phys. Rev. Lett. **106**, 213902 (2011).
8. D. Chremmos, Z. Chen, D. N. Christodoulides, and N. K. Efremidis, Phys. Rev. A **85**, 023828 (2012).
9. F. Courvoisier, A. Mathis, L. Froehly, R. Giust, L. Furfaro, P. A. Lacourt, M. Jacquot and J. M. Dudley, Opt. Lett. **37**, 1736 (2012).
10. M. A. Alonso and M. A. Bandres, Opt. Lett. **37**, 5175 (2012).
11. M. Leutenegger, R. Rao, R. A. Leitgeb and T. Lasser, Opt. Express **14**, 011277 (2006).
12. E. Greenfield, I. Kaminer, and M. Segev, "Observation of Periodic Accelerating Beams," in *Frontiers in Optics 2012/Laser Science XXVIII*, OSA Technical Digest (online) (Optical Society of America, 2012), paper FW1A.2.
13. J. A. Davis, D. M. Cottrell, J. Campos, M. J. Yzuel, and I. Moreno, Appl. Opt. **38**, 5004 (1999).